%% file: main.tex
\documentclass[11pt,english]{article}
  \usepackage[T1]{fontenc}
  \usepackage{lmodern}
  \usepackage[latin9]{inputenc}
  \usepackage{geometry}
  \usepackage{amsthm}
  \usepackage{amsmath}
  \usepackage{amssymb}
  \usepackage{graphicx}
  \usepackage{setspace}
  \usepackage[font={small}]{caption}
  \usepackage{color}
  \usepackage{braket}
  \usepackage{enumerate}	
  \usepackage{microtype}
  \usepackage{fancyhdr}
  \usepackage{titling}
  \usepackage{todonotes}
  \usepackage[affil-it]{authblk}
  \usepackage[round]{natbib}
\newenvironment{keywords}%
   {\begin{trivlist}\item[]{\bfseries\sffamily Keywords:}\ }
   {\end{trivlist}}
   
\begin{document}
  \title{On the thermodynamical cost of some interpretations of quantum theory}
\author[1]{Carina E. A. Prunkl\thanks{\texttt{carina.prunkl@balliol.ox.ac.uk}}
   }    
  \author[2]{Christopher G. Timpson\thanks{\texttt{christopher.timpson@bnc.ox.ac.uk}}
  }
  \affil[1]{Balliol College\\University of Oxford}
  \affil[2]{Brasenose College, University of Oxford} 

\date{\today}
\maketitle 
  \begin{abstract} 
\noindent Recently, \cite{cabello_thermodynamical_2016} claim to have proven the existence of an empirically verifiable difference between two broad classes of quantum interpretations. On the basis of three seemingly uncontentious assumptions, (i) the possibility of randomly selected measurements, (ii) the finiteness of a quantum system's memory, and (iii) the validity of Landauer's principle, and further, by applying computational mechanics to quantum processes, the authors arrive at the conclusion that some quantum interpretations (including central realist interpretations) are associated with an excess heat cost and are thereby untenable---or at least---that they can be distinguished empirically from their competitors by measuring the heat produced. Here, we provide an explicit counterexample to this claim and demonstrate that their surprising result can be traced back to a lack of distinction between system and external agent. By drawing the distinction carefully, we show that the resulting heat cost is fully accounted for in the external agent, thereby restoring the tenability of the quantum interpretations in question.

\end{abstract}
\begin{keywords}
Quantum Interpretations, Computational Mechanics, Landauer's Principle, Thermodynamics
\end{keywords}

\newpage  

  \tableofcontents
  \onehalfspacing
  
\newpage  
  
  \section{Introduction}

For nearly a century, physicists and philosophers alike have puzzled over how to interpret quantum theory, unable to decide unambiguously between a variety of more or less promising candidates. In a recent publication, \cite{cabello_thermodynamical_2016} put forward an argument which seeks to demonstrate the existence of a real, physical---as opposed merely to a metaphysical---difference between various interpretations of quantum mechanics. Moreover, the authors assert that it is in principle possible to measure this difference experimentally. Their argument is based on methods derived from computational mechanics---a growing field that is concerned with the simulation and prediction of stochastic processes. Interestingly, when applied to certain physical processes, computational mechanics is able to provide us with thermodynamical limitations on these processes \citep{wiesner_information-theoretic_2012,garner_when_2015}.  Cabello et al.'s argument is a concrete, foundationally motivated, application of computational mechanics which suggests that there is a thermodynamical cost to bear for a subset of quantum interpretations: perhaps a pathological one.

The link between thermodynamics and computational mechanics can be understood as follows: depending on the complexity, i.e., randomness, of a pattern that is to be simulated, greater or fewer resources are needed in order either to create the pattern (as in writing it on a shuffled medium) or to predict its future, given observations of past data sequences. We can take the computational system we are interested in simulating to be a black box, with the only accessible empirical data being its input and output variables. It can then be proven that there exists a machine, called an $\epsilon$-machine, which is predictively optimal and uses the minimum resources, while simulating the input-output behaviour of the target system \citep{crutchfield_inferring_1989}. For some thermodynamic systems this method shows up the limitations for work extraction via physical processes. Given a resource-theoretic understanding of thermodynamics (i.e., an understanding which conceives thermodynamics primarily to be a theory about what tasks one can perform when furnished with certain resources\footnote{Modern accounts include \citep{horodecki_fundamental_2013,wallace_thermodynamics_2014,brandao_second_2015,gour_resource_2015}, however, the underlying idea that thermodynamics is to be understood relative to an agent and her means goes back to \cite{maxwell_theory_1871} (c.f. \cite{myrvold_statistical_2011}) and was later famously promoted by \cite{jaynes_gibbs_1965}.}), one might say that computational mechanics can be considered a useful tool for the task of understanding and enhancing the foundations of thermal physics.

Cabello et al. begin by dividing the set of quantum interpretations into two subsets, what they term \textit{Type I} and \textit{Type II} interpretations. They then argue that Type I interpretations are associated with a thermodynamical cost, rendering such interpretations highly problematic: either one of the (very plausible) three assumptions must be given up, or there exists a surprising heat generation which could be ruled in or out experimentally (and one would be surprised indeed if such heat generation were in fact to be found). If correct, this result would seem an outstanding breakthrough whose far reaching consequences might not only force us to abandon some of the most popular interpretations of quantum mechanics (Type I interpretations include such favourites as de Broglie--Bohm theory, Everett, and dynamical collapse theories such as GRW, for example) but would shake the foundations of our understanding of the relationship between scientific theories and the underlying ontic structure of the world. 

 
 Our prime concern in this paper is to assess Cabello et al.'s argument and the tenability of their conclusions. But we also have their example in mind as a test-case for the application of computational mechanics in pursuit of dividends in foundations of physics.
 
 We will begin with a brief outline of stochastic input-output processes before presenting the argument of \cite{cabello_thermodynamical_2016}, which applies this mathematical machinery to quantum systems. We will then analyse why Cabello et al.'s argument about the thermodynamical costs of some quantum interpretations fails, including offering a straightforward counterexample. We will show that the adumbrated heat cost is in fact not associated with the quantum system itself at all---it is not of quantum origin---and thus controversies over quantum interpretations are not germane to it, nor it to them. Rather, the heat cost arises with the external experimental setup stipulated by Cabello et al.

 \section{Computational Mechanics and the Foundations of Quantum Mechanics} 

We begin by introducing the most important aspects of stochastic input-output processes, as they form the backbone of the argument. More detailed discussions may be found in \citep{barnett_computational_2015,crutchfield_inferring_1989}. 

\subsection{Computational Mechanics: Input-Output Processes}

The goal behind modelling systems' behaviour by input-output processes is to find the minimal structural requirements that produce a particular statistical pattern. To do so, one works backwards from the statistics of experimental outputs to then find the minimal amount of resources needed in order to simulate output strings that are statistically indistinguishable from the actual experimental result.

More formally: A stochastic process $\overleftrightarrow{Y}$ is described as a bi-infinite one-dimensional chain $...,Y_{-1},Y_0,Y_1,...$ of discrete random variables $\{Y_t\}$ with values $\{y_t\}$, where $t$ is a discrete time parameter and the direction of the arrow above the random variable indicates whether the chain extends to the past (left arrow), the future (right arrow) or to past \textit{and} future (left-right arrow) infinity. The $\{y_t\}$ are the particular values the random variable takes at time $t$ and in our case we can think of them as the output values of an experiment performed on the system. For example, for a spin-measurement on a qubit---the kind of case with which Cabello et al. will be concerned---the outcome-types could be ``up'' and ``down'' for example, taken from the output alphabet $\mathcal{Y}=\{\text{``up'',``down''}\}$. If not only the output but also the input is stochastic (in our case, this will correspond to a choice of spin-measurement basis, which will be taken to be random) the effect of the input random variable on the future statistics needs to be taken into account as well. Such a process must then be modelled by a so-called stochastic \textit{input-output process}, $\overleftrightarrow{Y}\vert\overleftrightarrow{X}$, with input values $\{x_t\}$ from an alphabet $\mathcal{X}$. The whole input-output process may then be described as a collection of stochastic processes $\overleftrightarrow{Y}\vert\overleftrightarrow{X}\equiv\{\overleftrightarrow{Y}\vert\overleftrightarrow{x}\}_{\overleftrightarrow{x}\in\overleftrightarrow{\mathcal{X}}}$, where we take each process $\overleftrightarrow{Y}\vert\overleftrightarrow{x}$ to correspond to all possible output sequences $\overleftrightarrow{Y}$ that could arise from one particular input sequence $\overleftrightarrow{x}$, drawn from the set of all possible input sequences $\overleftrightarrow{\mathcal{X}}$. 
 
The \textit{probability distribution}\footnote{We consider only stationary probabilities, which means that the probabilities are time translation invariant.} over the set of all possible output sequences, given a particular input sequence, is then given by what is called the \textit{channel's distribution}:

\begin{equation}
\mathbf{P}(\overleftrightarrow{Y}\vert\overleftrightarrow{x})=\{\mathbf{P}(\overleftrightarrow{Y}\in\sigma\vert\overleftrightarrow{X}=\overleftrightarrow{x})\}_{\sigma \subseteq \overleftrightarrow{\mathcal{Y}},\overleftrightarrow{x}\in\overleftrightarrow{\mathcal{X}}}
\end{equation}

 The idea is now to divide the input-output sequences into \textit{pasts} and \textit{future} and furthermore to divide the various input-output pasts into sets that yield the same distribution over input-output futures.  Two input-output pasts $\overleftarrow{z}=(\overleftarrow{x},\overleftarrow{y})$ and $\overleftarrow{z}^\prime$ that yield the same future input-output conditional probabilities $P(\overrightarrow{Y}\vert\overrightarrow{X},\overleftarrow{Z}=\overleftarrow{z})=P(\overrightarrow{Y}\vert\overrightarrow{X},\overleftarrow{Z}=\overleftarrow{z}^\prime)$ are then said to belong to the same \textit{causal state} $s$. Denote the set of causal states $\mathcal{S}$. The ${\epsilon}$-map is then introduced as the  mapping $\epsilon: \mathcal{\mathbb{\overleftarrow{\mathcal{Z}}}}\rightarrow \mathbb{\mathcal{S}}$ from any input-output past onto its corresponding causal state \citep{barnett_computational_2015}. This map also induces a probability distribution over the causal states, which since the process is stationary and $\epsilon$ is time-independent, is called the process' \textit{stationary distribution}. The causal states contain all the relevant information for optimally predicting the future output statistics of the system and contain as much information as any of its input-output pasts. Here we take the input sequences to be uniformly distributed. The minimal amount of information needed to be stored in order to predict future outputs optimally is then given by the Shannon information $H(\mathcal{S})$ and is called the \textit{statistical complexity}. This also quantifies the amount of resources needed in order to model the system's future behaviour. 

 \subsection{Foundations: Division of Interpretations into two Groups}

\cite{cabello_thermodynamical_2016} seek to use the above machinery in combination with a few plausible assumptions to raise difficulties for a group of well-known quantum interpretations. Their approach is to divide the set of quantum interpretations into two broad classes, based on their respective takes on quantum probabilities: Type I interpretations are interpretations that regard probabilities as determined by ``intrinsic properties of the system'' \citep[p.1]{cabello_thermodynamical_2016}. These properties typically change post-measurement, depending on the choice of measurement performed on the quantum system. Examples which they mention of  Type I interpretations include de Broglie-Bohm theory \citep{bohm_suggested_1952,goldstein_bohmian_2001}, many worlds interpretations, e.g. \citep{everett_relative_1957,wallace_emergent_2012}, Ballentine's statistical interpretation \citep{ballentine_statistical_1970}, modal interpretations \citep{lombardi_modal_2016} and consistent histories, as well as GRW dynamical collapse theories \citep{ghirardi_unified_1986} and Spekkens' toy model \citep{spekkens_evidence_2007}.

Type II interpretations, on the other hand, comprise those interpretations which treat probabilities as ``relational properties between an observer and the system'' \citep[p.1]{cabello_thermodynamical_2016}. According to such interpretations, the quantum state corresponds to the ``experiences an observer has of the observed system''\citep[p.1]{cabello_thermodynamical_2016}. To the class of Type II interpretations, the authors assign, amongst others, the Copenhagen interpretation, Wheeler's view \citep{wheeler_quantum_2014}, relational interpretations \citep{rovelli_relational_1996} and QBist interpretations \citep{fuchs_quantum_2007}.
  
We note to begin that there are some significant problems associated with such a division of interpretations. First, some of the interpretations listed as Type I do not seem unambiguously to belong in either camp: the Everett interpretation, for instance, may well be taken to imply that probabilities ought to be regarded relationally as opposed to being intrinsic properties of a quantum system. After all, the Everettian could argue, the `bare quantum formalism' is fully deterministic at the fundamental level. The probabilities, by contrast, occur at an emergent level and justifications of the Born rule are typically given via decision-theoretic arguments \citep{wallace_emergent_2012}, very strongly suggesting that probabilities in Everett, if they need an interpretation at all, are best interpreted as concerning relational experiences of an agent within a given branch. 

The Type I/ Type II distinction also seems problematic with regard to $\Psi$-epistemic interpretations, in which the quantum state is taken to represent information about how a system is\footnote{For details on the $\Psi$-ontic/$\Psi$-epistemic distinction, see \cite{harrigan_einstein_2010}.}. For instance, the probabilities occurring in Spekkens' toy model---which allegedly belongs to the Type I camp---are purely epistemic and result from an incomplete knowledge about the underlying ontic state\footnote{In analogy with Jaynes' view of probabilities in classical statistical mechanics, where the probability distribution over the underlying ontic state equally represents ignorance \citep{jaynes_gibbs_1965}} \citep{spekkens_evidence_2007}, as opposed to being intrinsic properties of the system: they merely occur on the level of the observer whereas the underlying ontic evolution could be, for all we know, a deterministic one. It thereby seems somewhat misplaced to regard these probabilities as ``observer-independent'' \citep[p.1]{cabello_thermodynamical_2016}, in the sense of being part of the furniture of the world, a notion more commonly found in the context of propensity interpretations of probabilities \citep{popper_logic_2002}, but certainly not in an epistemic context.

We may try to make clearer sense of the Type I/ Type II distinction by phrasing it differently in order to capture what Cabello et al. seem to have in mind: namely a distinction between interpretations that either endorse or reject the notion of an underlying ontic physical state. 
What the interpretations listed under Type II then have in common is that they all roughly follow along Niels Bohr's line: There is no quantum world. At least, there is not a free-standing mind- and agent- independent one. Type II interpretations thereby operate on a completely different level from Type I interpretations, the latter of which consider quantum mechanics as ultimately describing a free-standing mind-independent objective reality. Drawing a distinction between these fundamentally different conceptions of scientific theory on the basis of their conception of probabilities, however, is misleading and unsatisfying. The probabilities are mere epiphytes on a deeper, more pressing issue about the nature of science. 

Consequently, if Cabello et al. are indeed correct and Type I interpretations could be ruled-out experimentally, then their result would have profound consequences: it undermines one of our core traditional conceptions of scientific practice, namely that scientific theories tell us what the world independent of us is like.


\subsection{Three Assumptions}  

As Cabello et al. explain, their argument concerning Type I interpretations rests on three assumptions, which we can state as follows:

\begin{enumerate}[(i)]
\item Which measurement is performed on a system is decided randomly, and in particular independently of the state of the system. 
\item A (finite dimensional) quantum system has a limited memory.
\item Landauer's principle is valid.
\end{enumerate}
 
They go on to suggest that for Type I interpretations, it follows from (i) that a system's \textit{intrinsic properties} typically ought to \textit{change}, depending on which measurement is performed on the system. When the authors speak of `limited memory', (ii), they seem to have in mind something like the following: when a system is being measured in some basis, it typically generates a new value of its intrinsic property or set of intrinsic properties, which will determine the quantum probabilities of the future measurements. Measurement in a different basis will force the system to change these values in order to comply with the correct quantum probabilities, thereby deleting---or perhaps rather overwriting---the previous values. Due to its limited memory capacity, the system cannot possess all the intrinsic values that will determine the probabilities of all possible measurement series. We may assume that by `intrinsic properties' Cabello et al. have in mind properties that determine the probabilities of future measurement outcomes as opposed to properties specifying pre-determined definite outcomes, the reason being that this latter interpretation of ``intrinsic properties'' would force us to adapt some sort of hidden variable interpretation, which, by definition is only a subgroup of the Type I class. We therefore take it that the former meaning of `intrinsic properties' is intended.
 
Preliminary assumption (iii) is the validity of Landauer's principle. As it is often stated, Landauer's principle asserts that the erasure of information in a system's information bearing degrees of freedom is accompanied by an increase of entropy in the non-information bearing degrees of freedom \citep{landauer_irreversibility_1961}. This increase of entropy will lead to the dissipation of $kT\ln 2$ of heat per deleted bit, where $T$ is taken by Cabello et. al. to be the temperature of the system (though more usually it is taken to be the temperature of the environment) and $k$ is, of course, the Boltzmann constant.


\subsection{Heat Dissipation between Successive Measurements}

We are now invited to consider a single qubit, on which an observer performs successive projective measurements in one of the two Pauli-bases, $\sigma_x$ or $\sigma_z$, chosen at random with equal probability (assumption (i)). Since the state of the qubit changes after non-orthogonal measurements, within a Type I framework the internal properties of the system must change too. Given that the system has limited memory (assumption (ii)), it needs to generate new values that determine its future behaviour and store them in its memory. To do so, the system will need---as they put it---to `erase information'. Applying Landauer's principle (assumption (iii)), \cite{cabello_thermodynamical_2016} finally conclude that during this erasure process, heat is dissipated into the environment.

To quantify the amount of information that needs to be erased, they model the system as a computational machine, a black box that generates output strings on the basis of some input and its internal memory. The optimal and minimal machine that is able to produce output strings that are statistically indistinguishable from the actual experimental outcomes, will be, as we mentioned earlier, an $\epsilon$-machine \citep{crutchfield_inferring_1989}. It maximises the mutual information between input-output past and output future, thus simulating the statistical process, and with minimal resources.

Applying the computational mechanical machinery to the qubit in question, we identify the input random variable $X_t$ with the choice of measurement basis, randomly selected from the alphabet $\mathcal{X}=\{\sigma_x,\sigma_z\}$. The output variable $Y_t$ represents the measurement results and can take values $\pm 1$. The causal state after the measurement is simply taken to be the respective quantum state, $s_o=\ket{0}$, $s_1=\ket{1}$, $s_+=\ket{+}$ or $s_-=\ket{-}$ (with, note, no specific view here taken on the ontology or otherwise, of the quantum state).

The machine has a probability $1/2$ of changing its causal state after a measurement. Hence, half of the time, it must update its internal properties, and Cabello et al maintain that this requires the erasure of information. They say \textit{``The average information that must be erased per measurement is the information contained in the causal state previous to the measurement, $S_{t-1}$, that is not contained in the causal state after the measurement, $S_t$.''} \citep[p.2]{cabello_thermodynamical_2016}. It should be noted that this formulation is perhaps somewhat misleading, as it suggests that \textit{a particular} causal state itself carries a certain amount of information. In fact, it is not \textit{the} causal state to which we assign an entropy, but it is instead the probability distribution over causal states which is associated with an entropy and thereby with a Shannon information. Such an average over causal states however is not a causal state itself. 

In any case, the amount of information that needs to be erased is equal to the \textit{conditional entropy} of

\begin{equation}
I_{erased}=H(S_{t-1}\vert X_t,Y_t,S_t).\label{erased}
\end{equation}
(See Appendix.)


In the given experiment, the probability distribution over the causal states is uniform, which allows us to only consider a particular causal state $s_0$, brought about by a particular measurement $\sigma_z$ in order to determine $I_{erased}$. Together with the fact that $H(Y_t\vert S_t)=0$, the average erased information for the case of a successively measured qubit is then calculated to be 

 \begin{equation}
 I_{erased}=H(S_{t-1}\vert \sigma_z,s_0)=-\sum_{s_j\in\mathcal{S}}P(S_{t-1}=s_j\vert \sigma_z,s_0)\log P(S_{t-1}=s_j\vert \sigma_z,s_0). \label{probabilities}
 \end{equation}

 The three possible causal states at time $t-1$ are $s_0,s_+$ and $s_-$, with transition probabilities $1/2$, $1/4$ and $1/4$.\footnote{The authors write at time $t$, but must have intended $t-1$.} The conditional entropy then turns out to be $I_{erased}=-\frac{1}{2}\log \frac{1}{2} - 2\cdot \frac{1}{4}\log \frac{1}{4}=\frac{3}{2}$ bits.
 
\cite{cabello_thermodynamical_2016} thereby conclude that once we accept assumptions (i)--(iii), it follows that the system on average must dissipate $\frac{3}{2}kT\ln2$ units of heat per measurement, if understood as belonging to Type I.\footnote{The authors furthermore generalise their result to N-outcome measurements with an associated heat generation which scales linearly with N. Thus sufficient measurements on a single system would produce as much heat as you like. This does not sound promising for Type I interpretations.} In principle, the above experiment could be implemented in a lab. From our previous observations of measurements on quantum systems, it is however safe to say that we would be very surprised indeed to observe any such heat dissipation. Cabello et al. thereby suggest that Type I interpretations are unlikely to be representative of the world, at least if their plausible assumptions (i)--(iii) hold. The above argument supposedly does not apply to Type II interpretations, however, as for these interpretations ``measurement outcomes are created randomly when the observables are measured, without any need to overwrite information in the system and therefore without the system dissipating heat due to Landauer's principle'' \citep[p.3]{cabello_thermodynamical_2016}. 

In the conclusion of their paper Cabello et al. do canvass the possibility that one or more of assumptions (i)--(iii) might be thought to fail instead of Type I interpretations being lumbered with an excess heat cost, and in particular they judge that the de Broglie--Bohm theory and the Everett interpretation should not havehe excess heat quantity attached to them, as both interpretations violate (ii)---the finite memory assumption. As they see it, in the de Broglie--Bohm case this is because the ontology includes a continuous field (the unitarily evolving wave function), and in the Everett case, because one has a splitting into a plurality of worlds. But it is at best obscure that either de Broglie--Bohm or Everett should be thought to violate the finite memory assumption. In the de Broglie--Bohm case, from the fact that the quantum state is taken to be real\footnote{Putting aside those views which would see it as law-like---nomological rather than ontological.\citep{duerr_bohmian_1995,duerr_quantum_2012}} it does not follow that a system has infinite memory capacity: nearly all the state is irrelevant to the time evolution of the definite physical quantities most of the time anyway due to decoherence, whilst a particle's motion is only guided by the value of the wavefunction assigned to the region immediately surrounding it in any case. And one might note that realist collapse theories such as GRW \textit{also} have a continuous field in them: what difference could it make to judgements of memory capacity whether the continuous field (sometimes) jumps around stochastically (GRW) or whether it instead evolves determinisically but most of it being irrelevant to the evolution of a particle (de Broglie--Bohm)? With regard to Everett: even if there is branching, each world would be one in which the finite memory condition held, so each world would be one in which the excess heat cost obtained. If anything, this would look worse, rather than better, for Everett.

In fact, Cabello et al. are quite right that there is no excess heat cost which arises for de Broglie--Bohm or for Everett, but this is not for the reason they allege (the possibility of infinite memory capacity). Rather it is, as we shall see, an instance of a general proposition: there is no excess heat cost for any Type I interpretation over a Type II interpretation.
 
\section{Limitations of Computational Mechanics}

Let us now analyse how Cabello et al. arrive at their surprising results. We will begin with a brief recapitulation of Landauer's principle before delivering an explicit counterexample to Cabello et al.'s claim and then identifying the shortcomings of their argument.

\subsection{Interlude: Landauer's Principle and Irreversibility}

A great deal in this argument hinges on the application of Landauer's principle, often taken to be the claim that the implementation of a logically irreversible operation is accompanied by a dissipation\footnote{Heat dissipation is generally taken to be thermodynamically irreversible.} of $kT\ln 2$ units of heat per bit into the environment. An operation is considered to be logically irreversible, if the output of the operation does not uniquely determine the input \citep{landauer_irreversibility_1961}. Often, the concept of information is invoked in order to describe this logical irreversibility. \cite{wiesner_information-theoretic_2012} write that ``logically irreversible operations forget information about the computational device's preceding logical state.'' [p.4060].

To appreciate why such characterisations of Landauer's principle in terms of `forgetting' can be misleading, however, we consider a logically irreversibly operation which can be implemented without any heat cost, the so-called RAND operation \citep{maroney_absence_2005}. RAND randomises the logical state of a bit, regardless of its input state. Physically, we may think of an implementation of RAND in the standard way of considering a molecule in a box, in which a partition is included. The molecule is originally on the left side (or right side) of the box and the whole system is in contact with a heat bath at temperature $T_{HB}$. Implementing RAND then simply requires one to remove the partition, wait for a sufficiently long time and then re-insert the partition. The operation is logically irreversible because the output (the randomly distributed molecule) does not uniquely determine the input (the molecule being in one of the two mutually exclusive states)---but obviously there was no heat exchange with the environment during this process. In a very naive sense, the RAND operation has `erased information', but this erasure of information has taken place at no heat cost. Landauer's principle therefore cannot simply be the statement that the implementation of a logically irreversible operation leads to the dissipation of heat into the environment.

The most prominent application of Landauer's principle is the so-called \textit{Landauer erasure} process, a resetting operation that maps the state of a randomised bit back to some pre-defined initial state. For the above described molecule-in-a-box scenario, such an erasure can be implemented by removing the partition and then pushing it isothermally in from one side of the box, until the particle is found once again certainly in the left (or right) side of the box. For this last step, work is performed on the system and heat is transferred into the surrounding heat bath\footnote{Similarly in the quantum case for a qubit, one needs to step-wise raise one of the two energy levels to infinity \citep{barnett_computational_2015}.}. This is done in a thermodynamically reversible fashion and therefore corresponds to a \textit{heat transfer} and not to a \textit{heat dissipation}. Because the system ends up in a pre-defined state, the Landauer erasure is distinct from the RAND operation described above, although both of them are logically irreversible. In general, whether or not a given logical operation can be performed in a thermodynamically reversible or irreversible depends on the choice of implementation. In principle therefore, any logical operation can be implemented in a thermodynamically reversible fashion. More details on this can be found in \citep{maroney_generalising_2009}, who's approach we follow closely in this section.  

What Landauer's principle does is provide us with a link between logical operations and the fundamental microdynamics. Properly put, it states that a logical transformation, reversible or irreversible, must be accompanied by a minimal average heat dissipation into the environment according to

 \begin{equation}
 \langle \Delta Q\rangle\geq -T_{HB}\Delta S,\label{eq:landauer}
 \end{equation}

where $\Delta S$ refers to the difference in von Neumann entropy between the two physical ensembles that correspond to the two logical states $\alpha$ and $\beta$, and $\Delta Q$ is the heat generated in the environment \citep{maroney_generalising_2009}. Phrased like this, Landauer's principle not only becomes utterly unmysterious, it also becomes clear that whether or not heat is transferred into the environment solely depends on the von Neumann entropy of the system before and after the operation. This means in particular that one does not need to make any reference to information erasure or the like. 

If we re-write Landauer's principle in a way that makes use of the concept of information by having it explicitly include the entropy of the information bearing degrees of freedom, the change in (Gibbs-) von Neumann entropy is related to the change in Shannon entropy by $\Delta S=\sum_\beta P(\beta)S_{\beta}+\sum_\alpha P(\alpha)S_\alpha+k\Delta H \ln2$, with $\Delta H$ being the change in Shannon entropy. Making use of Equation (\ref{eq:landauer}), Landauer's principle then becomes 

\begin{equation}
\Delta S_{NI}\geq -k\Delta H\ln 2,\label{eq:landauerinfo}
\end{equation}

where the entropy change of the non-information bearing degrees of freedom is given by the entropy change in the environment and the weighted entropy changes of the sub-ensembles: $\Delta S_{NI}=\Delta S_{E}+\sum_\beta P(\beta)S_{\beta}-\sum_\alpha P(\alpha)S_{\alpha}$. This quantity is increasing for logically irreversible, deterministic operations. For non-deterministic operations however, it is decreasing, making Equation (\ref{eq:landauer}) more useful in the general context. 

Now: Let us consider applying Landauer's principle in the form of Equation (\ref{eq:landauer}) to the repeatedly measured quantum system. It is clear that once the measurement process is up and running, the quantum system will be in a maximally mixed state at each time step, independent of the chosen measurement basis. This means that the difference in the von Neumann entropy of the quantum system before and after each measurement is zero. Given, furthermore, that the density matrix is the appropriate entity for calculating the entropy of a quantum system---even if one's interpretation involves further variables as is the case in the de Broglie--Bohm theory for example---it follows that Landauer's principle does not predict a heat cost for Type I interpretations: the lower bound on heat dissipation into the environment is zero.\footnote{N.B. In the standard case in de Broglie--Bohm in which the distribution of particle positions is given by the Born rule then one will calculate the entropy of the system via its density matrix. If the distribution is not given by the Born-rule---the system is not in \textit{quantum equilibrium}, in the phrase---there will be rather more pressing departures from standard quantum predictions than simply thermal ones. \citep{valentini_signal-locality_1991,valentini_signal-locality_1991-1,valentini_subquantum_2002}} Notice that we have had to say nothing here of the storing or deleting or erasure of information.

But now, interestingly, we seem to have arrived at a contradiction: On the one hand, applying Landauer's principle to the successive measurements on a quantum system seems to entail a non-zero lower bound to the average heat cost per measurement. Whilst on the other it entails a zero lower bound. What needs to give?

In fact, nothing. This appearance of contradiction is misleading. The two results are not in fact in conflict. Both involve licit applications of Landauer's principle, but it is only the second (the zero heat minimum heat cost claim) which pertains to features of the quantum system.  

To explain this point and to help clarify where Cabello et al.'s argument has gone wrong, we will now construct an explicit counterexample to their argument. It will become apparent that the heat cost they calculate does not stem from the quantum system itself but from the particular setup that is chosen. The heat cost will be shown to be due to \textit{external} matters.

\subsection{A Counterexample: Type I without Heat Dissipation}

Spekkens' toy model \citep{spekkens_evidence_2007} is explicitly taken to be a Type I interpretation. We will make use of this simple and transparent framework in order to illustrate how successive measurements in non-orthogonal bases do \textit{not} lead to a predicted heat cost. 

In Spekkens' toy model, measurements of the system only yield incomplete knowledge about the underlying state in such a way that the maximal amount of knowledge about the ontic state equals the lack of knowledge about it. A Spekkens qubit can be in one of four possible ontic states, `1', `2', `3' or `4'. Measurements are identified with questions of the form: is the system in state 1 $\vee$ 2 or 3 $\vee$ 4 (measurement in the $\ket{0}/\ket{1}$ basis), or, alternatively, 1 $\vee$ 3 or 2 $\vee$ 4 (measurement in the $\ket{+}/\ket{-}$ basis)? A measurement result of $\ket{0}$ will therefore yield a state of knowledge of 1 $\vee$ 2. Impressively, quantum behaviour of a single qubit for measurements restricted to the $x$, $y$, and $z$ bases is fully recovered in this model.

If a system is prepared to be in 1 $\vee$ 3 but then measured to be in 1 $\vee$ 2, thereby performing the Spekkens equivalent of a non-orthogonal measurement, we know for a fact that the system must have been in ontic state 1 \textit{before} the measurement \citep{spekkens_evidence_2007}. This means that it is possible to know the precise ontic state of the system at an earlier time, but impossible to know it at the current time. A measurement in a non-orthogonal basis therefore \textit{disturbs} the system in such a way that its values become non-definite in its previous basis.

\begin{figure}[htb]
  \centering
  \def\svgwidth{\columnwidth}
    \resizebox{0.85\textwidth}{!}{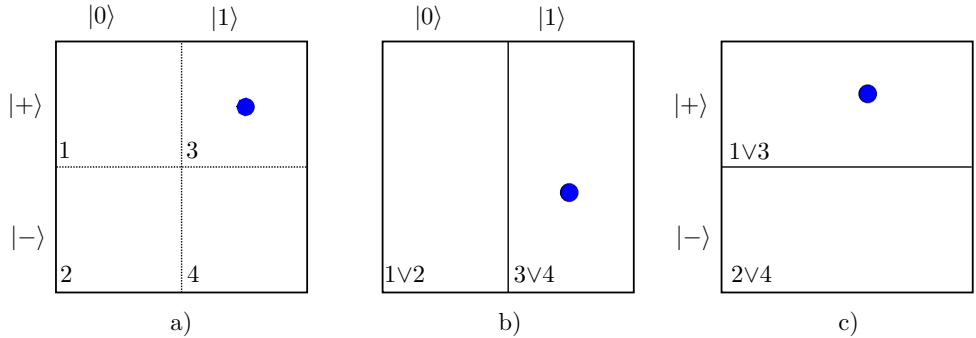}
  \caption{Illustration of standard measurements in Spekkens' toy model. We consider a classical particle in a two dimensional box. a) illustrates how at each instant, the particle is in one of 4 mutually exclusive ontic states. Measurements in the standard bases are modeled as the b) vertical or c) horizontal insertions of a partition into the box, followed by a location measurement that allows one to identify the particle's position as either being left/right or top/bottom. Two consecutive left/right and top/bottom measurements are non-commutative.}
  \label{fig:toymodel}
\end{figure}

Since we only consider measurements in the $x$ and $z$ directions, we can illustrate Spekkens' toy model conveniently by considering a classical particle entrapped in a two dimensional box in contact with a heat bath, as illustrated in Figure (\ref{fig:toymodel}a). 
Each section of the box corresponds to one of the four ontic states the particle can be in. To perform measurements, either a horizontal or a vertical partition can be inserted. A measurement in the $\ket{0}/\ket{1}$ basis for example corresponds to inserting a partition \textit{vertically} (Figure \ref{fig:toymodel}b) ) and measuring on which side of the partition the particle is found. Performing a measurement in the $\ket{+}/\ket{-}$ basis in contrast corresponds to the insertion of the partition \textit{horizontally} before measuring the the particle's position (Figure \ref{fig:toymodel}c) ). There can only be one partition in the box at a given time, and so in order to perform a measurement in the $\ket{+}/\ket{-}$ basis by inserting the partition horizontally, we need to remove the partition from the previous measurement in the $\ket{0}/\ket{1}$ basis. This must happen sufficiently quickly compared to the free motion of the particle so as to ensure that two consecutive measurements in the same basis yield the same result.

With the above setup at hand, we now implement Cabello et al.'s experiment and include successive random measurements in non-orthogonal bases. 

We begin by noting the need to distinguish between two notions of measurement which one might have in mind. On the one hand, one can consider some physical process taking place which leaves the system in some definite value of some observable. (A definite possessed value of the observable, whether or not anyone knows it.) On the other hand, one might consider an external observer or agent who comes to know what the definite value (or values) a system possesses at a given time are. (This difference is akin to the selective/non-selective measurement distinction familiar in quantum foundations.) Such an external observer need not be a person, but could be anything that reliably correlates with the measurement outcome, such as a memory cell.

We first consider the case in which there is no external agent. Measurements on the Spekkens particle are performed by either the horizontal or vertical insertion of the partition. In order to implement Cabello et al.'s experimental setup, we require that the orientation of the partition change randomly, in such a way that it has a probability of $50$\% of remaining in its previous position and a probability of $50$\% of changing its orientation from horizontal to vertical or vice versa. This is an implementation of Cabello et al.'s random variable $X_t$. At each time step, the system will be in a well defined state with definite values. It is evident \textit{prima facie} that there is no heat exchange with the environment at any point\footnote{Whether or not the act of removing and re-inserting the partition is thermodynamically reversible or irreversible is debatable given to the single particle nature of the experiment. One may argue that the removal of the partition resembles a free expansion, which is thermodynamically irreversible. However, if we take thermodynamic reversibility to be equal with the claim that the (Gibbs--) von Neumann entropy of the system remains constant, then the removal and re-insertion of the partition is indeed reversible, in accordance with \cite{maroney_generalising_2009}. Either way, there is clearly no heat exchanged with the environment.}. Moreover, this Spekkens setup is merely a more elaborate version of the previously introduced RAND operation. Evidently the system is in a definite state with definite values after each measurement, however, this ``generation'' of new values is not accompanied by a heat exchange with the environment.

One may object to the above reasoning by demanding that the partition \textit{itself} must have a reset state, that it therefore must delete information and reset itself after each measurement. However, the described Spekkens' setup does not require a costly \textit{Landauer erasure}: the system \textit{overwrites} its previous state after each time step\footnote{The behaviour of the partition for example could be implemented by a beam splitter, followed by a NOT operation on the one hand and the identity operation on the other hand. In particular, no reference to the partition's previous state must be made for either of those operations.}. As there is no external agent who records the measurement outcomes by correlating a measurement apparatus with the system's state, there are no resources needed for this implementation of measurements.

\begin{figure}[htb]
  \centering
  \def\svgwidth{\columnwidth}
    \resizebox{0.7\textwidth}{!}{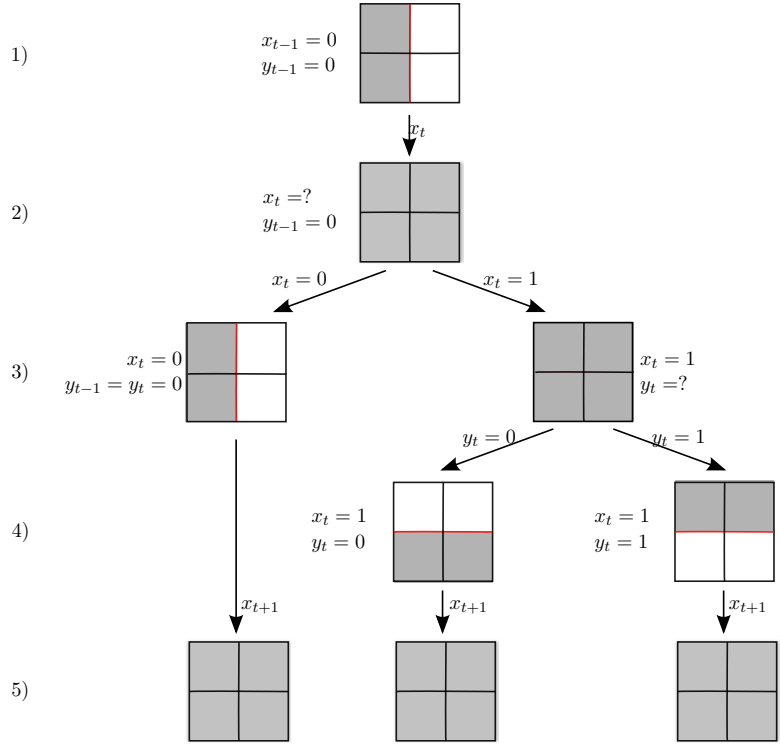}
  \caption{Illustration of the various steps in an exemplary measurement cycle.}
  \label{fig:memory}
\end{figure}

The situation changes if we require the measurement performed and the result obtained to be determined and/or recorded by an external agent. Such an agent need not be human, but could be any system that is able to correlate itself with the Spekkens system and perform operations depending on the outcome. In the case of the experimental setup described above, a (memoryless) external agent needs to acquire at least two bits of information in order to determine which state the system is in at a given time $t$: one bit that determines whether the particle was measured in the $x$ or $z$ basis and one bit that determines the outcome of the measurement, namely whether the particle is left/right or top/bottom respectively. We imagine now a situation in which the agent is memoryless, i.e. has no access to the choice of measurement basis and the measurement result at time $t-1$. If we allow the agent to have only two binary memory cells, one that reads out the choice of basis given by $X_t$, and one that reads out the position of the particle, $Y_t$, then once the agent has determined the state of the system at time $t$, she needs to reset both memory cells in order to prepare herself for the next measurement cycle at time $t+1$. Given that resetting memory cells is costly, there will be a heat cost of $kT\ln2$ associated with each measurement cycle.

This is where Cabello et al.'s result enters the picture. The described heat cost can in fact be reduced to $3/2$ bits once we allow the agent to use a recording device, or a \textit{memory}, which allows her to access the measurement results at the previous time-step\footnote{In practice, we need to grant the agent at least two more memory cells as resources, such as to be used as a memory for the measurement basis and measurement result at time $t-1$. In computational mechanics it is more common simply to supply the agent with an empty tape on which she records the measurement outcomes and at the same time allow her to access the tape on which the basis choices are written, thereby providing her with $\overleftarrow{y_t}$ and $\overleftarrow{x_t}$.} $t-1$. In this case, the agent does not need to perform the position measurement iff $x_t=x_{t-1}$, i.e. if she finds that the measurement basis at time $t$ is the same as at time $t-1$. This is due to the fact that consecutive measurements in the same basis always yield the same measurement results. Given that she has access to the previous measurement result, she can therefore skip the location measurement in $50$\% of the cases and hence save resources. If the agent has access to a memory that specifies the previous measurement basis and measurement outcome, the average amount of bits needed to specify the measurement and outcome at time $t$ can thereby be reduced to $3/2$ bits. The minimal average heat cost per measurement cycle therefore becomes $\frac{3}{2} k T\ln2$, the value Cabello et al. derived.

Figure \ref{fig:memory} illustrates the various steps in an exemplary measurement cycle: after an input $x_t$, the state of the system has changed into a maximally mixed state. The first step of the agent must be to determine the measurement basis, i.e. the position of the partition. This step must be performed during \textit{each} cycle, and so, given a finite memory for the agent and the need to create new blank memory states, there is always a heat cost of $kT\ln 2$ associated with this step. If $x_t=x_{t-1}$, nothing further happens, $y_{t-1}$ changes into $y_t$, but no resources are required for this step and so the cycle is finished. If $x_t\neq x_{t-1}$ however, a position measurement must be performed that determines $y_{t}$. This once more leads to $kT\ln 2$ units of heat, since the agent only has limited memory and therefore needs to reset her memory before each measurement. Since this measurement of $y_{t}$ is only performed half of the time, the average heat cost associated with it is only $\frac{1}{2} kT\ln2$ units of heat. Adding up the various contributions leads to an average heat cost of $\frac{3}{2} kT\ln 2$ per measurement cycle, in accordance with Cabello et al.'s results. From a global point of view, however, the Spekkens quantum system itself is in a maximally mixed state from step 2) onwards.

The alleged quantum heat cost therefore merely results from the need to reset the various memory cells of the agent that are needed to determine the measurement outcome, making its existence no more mysterious than the heat cost involved in the consecutive measurement of the outcomes of a fair coin flip, given limited resources: at each time step the measurement apparatus must be reset so as to be able to perform the consecutive measurements. Differently put, in terms of particles and boxes: performing consecutive RAND operations on a system leads to a heat cost iff the system's state is recorded at each time step, in which case the measurement apparatus needs to be reset before each measurement. What Cabello et al. have shown therefore, is simply that if we allow the agent to have a \textit{memory}, the average heat cost for repeatedly recording the outcomes can be reduced from $2$ bits to $3/2$ bits, and no further.

Furthermore, in order to explain the origin of this heat cost, we did not need to mention any \textit{intrinsic values} which supposedly determine the future behaviour of the quantum system. The resources needed for Cabello et al.'s experiment are resources that are to be provided by the observer who wants to record the measurement results, and not by the quantum system itself. We note that the observer and his or her resources could well be purely classical; and we note moreover that the very same heat cost would be incurred even for a Type II interpretation.

To emphasize how talk about `value generation' is misplaced, we finally consider one last example of an input-output process, which equally profits from the presence of an external memory: a feedback NOR channel, where the current output is determined by the input and the previous output in the standard NOR fashion. Such a channel has two causal states, and we take it to be driven by a random input $X_t$, just like before.  Calculating the entropy difference between two consecutive time steps results, similarly to Cabello et al.'s result, in $I^*_{erased}=H(Y_{t-1}\vert X_t Y_t)=1/2$ bits. Given that this channel is fully deterministic and only driven by the random variable $X_t$, it is evident that any talk about the generation of values that determine the system's future behaviour is misplaced.

\subsection{Some Remarks on Computational Mechanics}

The above examples demonstrate that the heat cost attributed to certain types of quantum interpretation by Cabello et al. leads back to an \textit{external} agent repeatedly performing measurements with limited information storage available. The external agent does not need to be human but could equally be a machine and the calculated heat cost indeed gives a lower bound to the energy required to run such process. Cabello et al.'s result therefore clearly plays an important role for determining the resources  needed to perform certain quantum computational tasks.  What their argument does not establish however, is a distinctive and perhaps problematic heat cost for Type I interpretations. We can trace back the misinterpretation of their mathematical results to a lack of discrimination between the agent who acts and the system that is acted upon. Instead, agent and system together were treated as a unified computational machine, leading to the erroneous conclusion that the individual system itself was driving the computation and was the locus of the heat cost. This suggests that we need to be more careful when we apply computational mechanics to the foundations of quantum mechanics. In particular the tempting but somewhat misleading idea that information is a \textit{concrete} rather than \textit{abstract} entity, which gives it the status of a physical substance that can be transferred and destroyed, leads to intuitions that must be double checked by contrasting them with real, physical situations (cf. \citep{timpson_quantum_2013} for more details on the concrete vs. abstract distinction).

\section{Conclusion}

We have analysed the claim that there is a physical difference between two classes of quantum interpretations in terms of an excess heat generation in successive measurements, a difference which could in principle be tested experimentally. This is not so: there is no such differential heat production. In so far as there is a heat cost associated with successive randomly selected measurements on a quantum system, it arises from accounting for the external resources needed to record the performance and results of the measurement operations on the system, and not from a putative erasure process that takes place within the quantum system itself. That there is no heat cost involved with consecutive measurements of this kind arising from the quantum system itself---regardless of how one interprets probabilities---can immediately be seen upon calculating the difference of the von Neumann entropy of the quantum system between the various time steps, which difference is zero. This fact is reconciled with Cabello et al.'s quite correct mathematical result precisely by noting that the latter only concerns---when properly understood---costs of the external record, of the external agent.

The question of which interpretations might truly represent our world thereby remains unanswered by thermodynamical considerations concerning successive measurement scenarios.

\section*{Acknowledgements}
We would like to thank Adan Cabello, Mile Gu, Andrew Garner and Vlatko Vedral for helpful discussions as well as David Wallace and Harvey Brown for useful comments. This work was partially supported by a grant from the Templeton World Charity Foundation and the British Society for the Philosophy of Science.

  \newpage
  \appendix 
  \section*{Appendix}

The quantity $I_{erased}$ has its origin in the difference of the Shannon entropies at times $t$ and $t-1$ as can be seen in the following, short derivation.

\begin{align}
H\left(X_tY_tS_t\right)-H\left(X_{t}Y_{t-1}S_{t-1}\right) &=H\left(X_tY_tS_{t-1}\right)-H\left(X_tY_{t-1}S_{t-1}\right)-H\left(S_{t-1}\vert X_t Y_t S_{t}\right) \nonumber \\
&= H\left(X_t\right)+H\left(Y_t\vert S_t\right)+H\left(S_t\right) \nonumber\\
&-H\left(X_t\right)-H\left(Y_{t-1}\vert S_{t-1}\right)-H(S_{t-1})-H\left(S_{t-1}\vert X_tY_tS_t\right) \nonumber \\
&=-H\left(S_{t-1}\vert X_t Y_t S_t\right) = - I_{erased} \nonumber,
\end{align}
where we assumed that the process is \textit{unifilar}, namely $H\left(S_t\vert X_tY_tS_{t-1}\right)=0$, that the input is time-independent, i.e. $H\left(X_{t-1}\right)=H\left(X_t\right)$ and that the current causal state determines the output uniquely $H\left(Y_t\vert S_t\right)=0$. Following Landauer's Principle, the physical implementation of a logical operation leads to a generation of heat equal to at least $-kT\ln 2$ times the total change in Shannon entropy. To calculate the total change, all involved random variables at a given time need to be taken into account. $I_{erased}$ therefore provides a lower bound for the heat cost involved in the implementation of Cabello et al.'s experiment.

\newpage

  \bibliography{bibliography}
  \end{document}

%% file: toymodel.pdf_tex
\begingroup%
  \makeatletter%
  \providecommand\color[2][]{%
    \errmessage{(Inkscape) Color is used for the text in Inkscape, but the package 'color.sty' is not loaded}%
    \renewcommand\color[2][]{}%
  }%
  \providecommand\transparent[1]{%
    \errmessage{(Inkscape) Transparency is used (non-zero) for the text in Inkscape, but the package 'transparent.sty' is not loaded}%
    \renewcommand\transparent[1]{}%
  }%
  \providecommand\rotatebox[2]{#2}%
  \ifx\svgwidth\undefined%
    \setlength{\unitlength}{553.37678375bp}%
    \ifx\svgscale\undefined%
      \relax%
    \else%
      \setlength{\unitlength}{\unitlength * \real{\svgscale}}%
    \fi%
  \else%
    \setlength{\unitlength}{\svgwidth}%
  \fi%
  \global\let\svgwidth\undefined%
  \global\let\svgscale\undefined%
  \makeatother%
  \begin{picture}(1,0.34129257)%
    \put(0,0){\includegraphics[width=\unitlength]{toymodel.pdf}}%
    \put(0.08141895,0.32321041){\color[rgb]{0,0,0}\makebox(0,0)[lb]{\smash{$\ket{0}$}}}%
    \put(0.20921065,0.32156145){\color[rgb]{0,0,0}\makebox(0,0)[lb]{\smash{$\ket{1}$}}}%
    \put(-0.00082448,0.23131036){\color[rgb]{0,0,0}\makebox(0,0)[lb]{\smash{$\ket{+}$}}}%
    \put(0.00186034,0.09417829){\color[rgb]{0,0,0}\makebox(0,0)[lb]{\smash{$\ket{-}$}}}%
    \put(0.42034856,0.32283259){\color[rgb]{0,0,0}\makebox(0,0)[lb]{\smash{$\ket{0}$}}}%
    \put(0.54814027,0.32118366){\color[rgb]{0,0,0}\makebox(0,0)[lb]{\smash{$\ket{1}$}}}%
    \put(0.38914004,0.05493545){\color[rgb]{0,0,0}\makebox(0,0)[lb]{\smash{1$\lor$2}}}%
    \put(0.5233808,0.0552291){\color[rgb]{0,0,0}\makebox(0,0)[lb]{\smash{3$\lor$4}}}%
    \put(0.05167946,0.18339352){\color[rgb]{0,0,0}\makebox(0,0)[lb]{\smash{1}}}%
    \put(0.05313319,0.05493545){\color[rgb]{0,0,0}\makebox(0,0)[lb]{\smash{2}}}%
    \put(0.18386303,0.18313536){\color[rgb]{0,0,0}\makebox(0,0)[lb]{\smash{3}}}%
    \put(0.68878414,0.23131036){\color[rgb]{0,0,0}\makebox(0,0)[lb]{\smash{$\ket{+}$}}}%
    \put(0.69146892,0.09417826){\color[rgb]{0,0,0}\makebox(0,0)[lb]{\smash{$\ket{-}$}}}%
    \put(0.74618978,0.18254967){\color[rgb]{0,0,0}\makebox(0,0)[lb]{\smash{1$\lor$3}}}%
    \put(0.74866808,0.05493545){\color[rgb]{0,0,0}\makebox(0,0)[lb]{\smash{2$\lor$4}}}%
    \put(0.16774608,0.00486784){\color[rgb]{0,0,0}\makebox(0,0)[lb]{\smash{a)}}}%
    \put(0.50757135,0.00486784){\color[rgb]{0,0,0}\makebox(0,0)[lb]{\smash{b)}}}%
    \put(0.86005241,0.00486784){\color[rgb]{0,0,0}\makebox(0,0)[lb]{\smash{c)}}}%
    \put(0.18471575,0.05493545){\color[rgb]{0,0,0}\makebox(0,0)[lb]{\smash{4}}}%
  \end{picture}%
\endgroup%

%% file: memory.pdf_tex
\begingroup%
  \makeatletter%
  \providecommand\color[2][]{%
    \errmessage{(Inkscape) Color is used for the text in Inkscape, but the package 'color.sty' is not loaded}%
    \renewcommand\color[2][]{}%
  }%
  \providecommand\transparent[1]{%
    \errmessage{(Inkscape) Transparency is used (non-zero) for the text in Inkscape, but the package 'transparent.sty' is not loaded}%
    \renewcommand\transparent[1]{}%
  }%
  \providecommand\rotatebox[2]{#2}%
  \ifx\svgwidth\undefined%
    \setlength{\unitlength}{393.29759549bp}%
    \ifx\svgscale\undefined%
      \relax%
    \else%
      \setlength{\unitlength}{\unitlength * \real{\svgscale}}%
    \fi%
  \else%
    \setlength{\unitlength}{\svgwidth}%
  \fi%
  \global\let\svgwidth\undefined%
  \global\let\svgscale\undefined%
  \makeatother%
  \begin{picture}(1,0.92790248)%
    \put(0,0){\includegraphics[width=\unitlength]{memory.pdf}}%
    \put(0.49880599,0.76657628){\color[rgb]{0,0,0}\makebox(0,0)[lb]{\smash{$x_t$}}}%
    \put(0.32707737,0.57607209){\color[rgb]{0,0,0}\makebox(0,0)[lb]{\smash{$x_t=0$}}}%
    \put(0.59328055,0.57607209){\color[rgb]{0,0,0}\makebox(0,0)[lb]{\smash{$x_t=1$}}}%
    \put(0.56696901,0.37765991){\color[rgb]{0,0,0}\makebox(0,0)[lb]{\smash{$y_t=0$}}}%
    \put(0.80550765,0.37765991){\color[rgb]{0,0,0}\makebox(0,0)[lb]{\smash{$y_t=1$}}}%
    \put(0.28856797,0.16828243){\color[rgb]{0,0,0}\makebox(0,0)[lb]{\smash{$x_{t+1}$}}}%
    \put(0.55092403,0.16828243){\color[rgb]{0,0,0}\makebox(0,0)[lb]{\smash{$x_{t+1}$}}}%
    \put(0.90240743,0.16828243){\color[rgb]{0,0,0}\makebox(0,0)[lb]{\smash{$x_{t+1}$}}}%
    \put(-0.00062842,0.05858733){\color[rgb]{0,0,0}\makebox(0,0)[lb]{\smash{5)}}}%
    \put(-0.00048689,0.65927852){\color[rgb]{0,0,0}\makebox(0,0)[lb]{\smash{2)}}}%
    \put(0.3158432,0.69707242){\color[rgb]{0,0,0}\makebox(0,0)[lt]{\begin{minipage}{0.1452697\unitlength}\raggedright $x_{t}=?$ $y_{t-1}=0$\end{minipage}}}%
    \put(-0.0014097,0.85891506){\color[rgb]{0,0,0}\makebox(0,0)[lb]{\smash{1)}}}%
    \put(0.31401505,0.8934837){\color[rgb]{0,0,0}\makebox(0,0)[lt]{\begin{minipage}{0.15344133\unitlength}\raggedright $x_{t-1}=0$ $y_{t-1}=0$\end{minipage}}}%
    \put(-0.0002944,0.25866927){\color[rgb]{0,0,0}\makebox(0,0)[lb]{\smash{4)}}}%
    \put(0.37702558,0.29333569){\color[rgb]{0,0,0}\makebox(0,0)[lt]{\begin{minipage}{0.12944825\unitlength}\raggedright $x_{t}=1$ $y_t=0$\end{minipage}}}%
    \put(0.73341825,0.29333569){\color[rgb]{0,0,0}\makebox(0,0)[lt]{\begin{minipage}{0.12944825\unitlength}\raggedright $x_{t}=1$ $y_t=1$\end{minipage}}}%
    \put(-0.00063407,0.45884476){\color[rgb]{0,0,0}\makebox(0,0)[lb]{\smash{3)}}}%
    \put(0.06424867,0.4921042){\color[rgb]{0,0,0}\makebox(0,0)[lt]{\begin{minipage}{0.14958464\unitlength}\raggedleft $x_{t}=0$ $y_{t-1}=y_t=0$\end{minipage}}}%
    \put(0.79082046,0.49206187){\color[rgb]{0,0,0}\makebox(0,0)[lt]{\begin{minipage}{0.12944825\unitlength}\raggedright $x_{t}=1$ $y_t=?$\end{minipage}}}%
  \end{picture}%
\endgroup%